\documentclass[acmtog, nonacm]{acmart}

\usepackage{arydshln}
\usepackage[nolist,nohyperlinks]{acronym}
\acrodef{cvae}[CVAE]{Conditional Variational Autoencoder}
\acrodef{vae}[VAE]{Variational Autoencoder}
\acrodef{fvae}[Flow-VAE]{Flow Variational Autoencoder}
\acrodef{mgl}[MoGlow]{Motion Glow}
\acrodef{lstm}[LSTM]{Long Short Term Memory}
\acrodef{blstm}[BLSTM]{Bi-Directional Long Short Term Memory}
\acrodef{ohe}[OHE]{One Hot Encoding}
\acrodef{elbo}[ELBO]{Evidence Lower Bound}
\acrodef{mse}[MSE]{Mean Squared Error}
\acrodef{kld}[$D_{KL}$]{Kullback–Leibler Divergence}
\acrodef{relu}[ReLU]{Rectified Linear Unit}
\acrodef{gan}[GAN]{Generative Adversarial Network}
\acrodef{rnn}[RNN]{Recurrent Neural Network}
\acrodef{nf}[NF]{Normalising Flow}
\acrodef{anv}[ANOVA]{Analysis of Variance}
\acrodef{mfcc}[MFCC]{Mel-frequency Cepstral Coefficient}
\acrodef{bert}[BERT]{Bidirectional Encoder Representations from Transformer}
\acrodef{tts}[TTS]{Text-To-Speech}
\acrodef{stt}[STT]{Speech-To-Text}
\acrodef{NMI}[NMI]{Normalised Mutual Information}
\acrodef{pase}[PASE+]{Problem Agnostic Speech Encoding}
\acrodef{bvh}[BVH]{Biovision hierarchical data}
\acrodef{genea}[GENEA]{Generation and Evaluation of Non-verbal Behaviour for Embodied Agent}
\acrodef{llm}[LLM]{Large-Language Model}
\acrodef{bpe}[BPE]{Byte-Pair Encoding}
\acrodef{sota}[SotA]{State-of-the-Art}
\acrodef{csmp}[CSMP-Diff]{Contrastive Speech and Motion Pretraining Diffusion}
\acrodef{clip}[CLIP]{Contrastive Language-Image Pre-Training}

\AtBeginDocument{%
  \providecommand\BibTeX{{%
    \normalfont B\kern-0.5em{\scshape i\kern-0.25em b}\kern-0.8em\TeX}}}

\newcommand{\llama}{\textsc{Llama}}
\newcommand{\llanimation}{\textsc{Lla}ni\textsc{ma}tion}

\begin{teaserfigure}
\centering
\includegraphics[width=\linewidth]{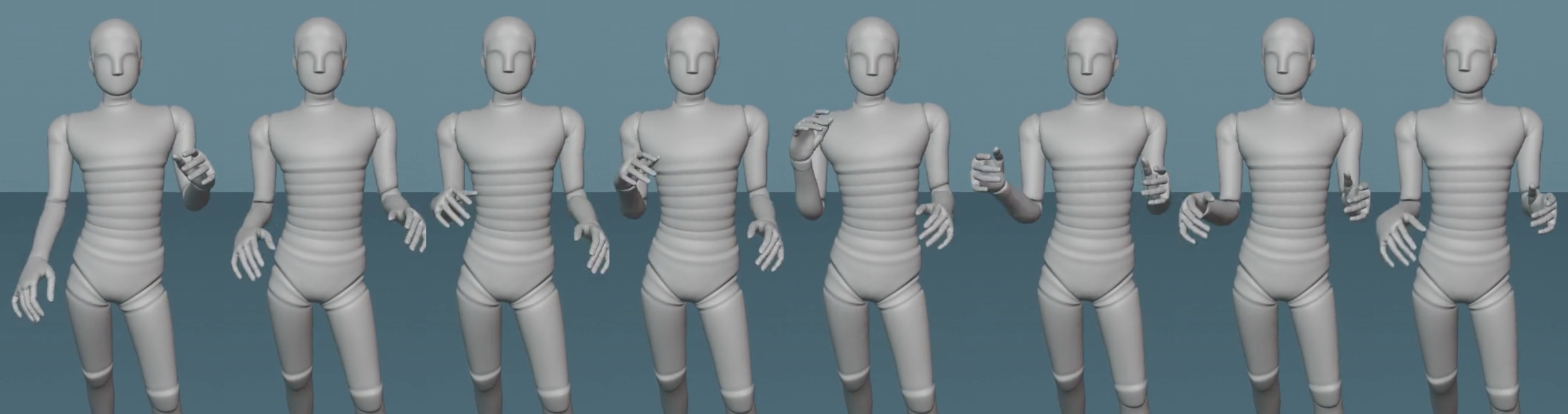}
\caption{We present a method for generating gestures from speech using \llama2 features derived \emph{from text} as the primary input, producing temporally aligned and contextually accurate gesture animation. This sequence generated from our method shows a speaker mimicking the use of a fork with their right hand while describing eating crab.}
\label{fig:teaser}
\end{teaserfigure}

\begin{document}

\title{\llanimation: {\llama} Driven Gesture Animation}


\author{Jonathan Windle}
\affiliation{%
  \institution{University of East Anglia}
  \country{UK}}

\author{Iain Matthews}
\affiliation{%
  \institution{University of East Anglia}
  \country{UK}}

\author{Sarah Taylor}
\affiliation{%
  \institution{Independent Researcher}
  \country{UK}}

\renewcommand{\shortauthors}{Windle et al.}

\begin{abstract}
Co-speech gesturing is an important modality in conversation, providing context and social cues. 
In character animation, appropriate and synchronised gestures add realism, and can make interactive agents more engaging. 
Historically, methods for automatically generating gestures were predominantly audio-driven, exploiting the prosodic and speech-related content that is encoded in the audio signal.
In this paper we instead experiment with using \ac{llm} features for gesture generation that are extracted from text using \llama2. 
We compare against audio features, and explore combining the two modalities in both objective tests and a user study. 
Surprisingly, our results show that \llama2 features on their own perform significantly better than audio features and that including both modalities yields no significant difference to using \llama2 features in isolation.
We demonstrate that the \llama2 based model can generate both beat and semantic gestures without any audio input, suggesting \ac{llm}s can provide rich encodings that are well suited for gesture generation. 

\end{abstract}



\received{20 February 2007}
\received[revised]{12 March 2009}
\received[accepted]{5 June 2009}

\maketitle

\section{Introduction}
Co-speech gesturing plays a crucial role in communication, as gestures effectively convey emotions and emphasis and enhance interactions by introducing social and contextual cues. These cues contribute to increased understanding, improved turn-taking, and enhanced listener feedback \cite{kendon1994gestures,mcneill1985so,studdert1994hand,de2012interplay}. 
Gestures have been found to influence what a listener hears ~\cite{bosker2021beat}, emphasising the importance of accurately depicting body motion during speech 
in entertainment applications such as video production, video games, avatars, virtual agents, and robotics. 
The ability to automatically produce realistic gestures from speech has broad applications in these areas.

There is a co-dependency between speech and gesture, where gesture production is a complex function of the speech content, semantics and prosody. 
For instance, beat gestures synchronise with the timing of the speech audio dynamics, while iconic gestures convey the shape of the discussed topic~\cite{bosker2021beat}.
Current research often focuses on speech-to-gesture generation using audio features as the primary modality.
While audio features are effective in encoding prosody, they may not capture semantics as well.
Conversely, text features capture the content, but may lack prosodic information.
It becomes apparent that a combination of features may yield optimal results.

\acfp{llm} are exposed to large natural language corpora, making them exceptional in language and content understanding. 
In this paper we explore the integration of \ac{llm} embeddings into a gesture generation model to improve the semantic accuracy of co-speech gestures. 
We experiment with methodologies for combining \ac{llm} embeddings with audio features, and report the objective and perceptual performance to determine the contribution of each feature.
Surprisingly, our results show that \ac{llm} features on their
own perform significantly better than audio features, and no significant difference is recorded when these two modalities are used in combination. 
Our approach, called \llanimation, utilises \llama2 language embeddings~\cite{llama2} and optionally combines them with \ac{pase} audio features~\cite{ravanelli2020multi} in a Transformer-XL architecture \cite{windlegenea23}. 
We show that our \ac{llm}-based {\llanimation} produces gestures that exhibit varied motions, capturing both beat and semantic gestures.
Our key contributions are as follows:
\begin{itemize}
    \item We are first to integrate \ac{llm} features into a gesture animation model.
    \item We evaluate the performance impact of using \llama2 features in combination with audio features and in isolation.
    \item We demonstrate that \ac{llm} features contribute more to the perceived quality of the resulting gesture animations than audio features.
\end{itemize}
\section{Related Work}
Speech-driven gesture generation has historically relied on audio features as its primary input. While text-based features have gained momentum in recent research, the utilisation of \ac{llm} features remains limited.
We review features used in gesture generation and on \ac{llm} models used in gesture and related fields.

\subsection{Speech Features for Gesture Generation}
Gesture generation systems widely adopt audio-based features. In the review by \cite{nyatsanga2023comprehensive} out of 40 methods reviewed, 35 used audio as an input feature. In contrast, only 17 methods involved text as an input.
Audio features can be embedded using various methods. Perhaps most common is the use of \ac{mfcc} \cite{hasegawa2018evaluation,alexanderson2020style,qian2021speech,habibie2022motion,pang2020cgvu, alexanderson2023listen}, sometimes combined with other prosodic features such as pitch (F0) and energy \cite{kucherenko2019analyzing}. Other latent representations such as Wav2Vec 2.0 \cite{baevski2020wav2vec} and \ac{pase} \cite{ravanelli2020multi} have grown in popularity as these can also effectively encode important speech-related information as well as prosodic features \cite{windle2022uea, windlegenea23,ng2024audio}, while improving speaker independence of the representation.
Audio features are advantageous with regard to beat gesture performance as these have a close relationship to prosodic activity, such as acoustic energy and pitch \cite{windle2022arm, pouw2020energy}.

Numerous approaches leverage a combination of both audio and text features, with different methods for incorporating textual information. Word rhythm was used by \cite{zhou2022gesturemaster} where words are encoded in a binary fashion, taking the value 1 if a word is spoken and 0 if not. Other works, such as those by \cite{windle2022uea,windlegenea23} and  \cite{yoon2020speech} integrate FastText embeddings and \cite{bojanowski2017enriching} which extend the Word2Vec approach \cite{word2vec} exploiting sub-word information.
BERT features \cite{devlin2018bert} have been successfully used in conjunction with audio in the work of \cite{ao2022rhythmic}. BERT, originally designed for language modelling and next-sentence prediction, is composed of transformer encoder layers. 

Using text as the exclusive input for gesture generation is infrequent, and performance is often limited when used.  \cite{yoon2019robots} and \cite{bhattacharya2021text2gestures} employ word embedding vectors \cite{pennington2014glove} to facilitate gesture generation. 

Despite the recognised advantages of text-based features, to the best of our knowledge, \ac{llm}s have not been used in the context of gesture generation, whether in isolation or in combination with audio inputs. This highlights a gap in the current research landscape that we aim to explore in this paper.

\subsection{Large Language Models}
Given the close relationship between language and gesture, the recent advances in \ac{llm} performance present a promising avenue for advancing gesture generation.
We provide a brief overview of \ac{llm} approaches and refer the reader to \cite{yang2023harnessing} for a comprehensive review.

\ac{llm} approaches fall into two categories: Encoder-Decoder/ Encoder only and Decoder only, often referred to as Bidirectional Encoder Representations from Transformers (BERT) \cite{devlin2018bert} and Generative Pre-trained Transformer (GPT) style, respectively. These models typically exhibit a task-agnostic architecture. 
Our primary focus in this work is on GPT-style models, which currently stand as leaders in \ac{llm} performance.

Numerous GPT-style models have been introduced, and among them, GPT-4 from \cite{openai2023gpt4} has emerged as a top performer across various language-based tasks. However, GPT-4 is a closed-source solution. 
The leading open-source alternative is currently \llama2 \cite{llama2}, which surpasses other open-source \ac{llm}s in tasks related to commonsense reasoning, world knowledge, and reading comprehension. 

\ac{llm}s have begun to garner attention in gesture-based tasks. For instance, \cite{hensel2023large} uses ChatGPT \cite{openai2023gpt4} for the selection and analysis of gestures, while \cite{zeng2023gesturegpt} uses ChatGPT to analyse and comprehend performed gestures.
To the best of our knowledge, there are no established methods for generating gestures directly from \ac{llm}s.
\section{Method}
In our exploration of using \ac{llm}s as a primary feature for co-speech gesture generation, we introduce \llanimation. \llanimation\space utilises \llama2 text embeddings, which can be used as an independent feature or in conjunction with \ac{pase} \cite{ravanelli2020multi} audio features. 
We are the first to integrate a \ac{llm} in this way. 
The generative model is based on the adapted Transformer-XL architecture presented by \cite{windlegenea23}.

\subsection{Speech Features}
Our method can leverage both audio and text-based features. Each modality has differing sample rates, with audio sample values updating at a faster pace than text tokens. We extract features at their original sample rates and align them to fit the timing of a motion frame at 30fps. We use $N$ to represent the number of $\approx$33ms motion frames in an input sequence. 
The PASE+ and \llama2 model weights are frozen and not updated during training. 

\subsubsection{Audio}
Audio features are extracted using the \ac{pase} model as these have been proven effective for gesture generation \cite{windle2022arm, windle2022uea, windlegenea23}. 
\ac{pase} was trained by solving 12 regression tasks to learn important speech characteristics using a multi-task learning approach. These tasks include estimating MFCCs, FBANKs and other speech-related information, including prosody and speech content. 
Using this model, we extract audio feature embeddings of size 768 for each 33ms audio window to align with the 30fps motion. Consequently, audio feature vectors, $A$, with a shape of $(N, 768)$ are generated for each audio clip.

\subsubsection{Text}
Word-level features are extracted using the pre-trained, 7-billion parameter \llama2 model \cite{llama2}. \llama2 adopts a Transformer architecture and has been trained on a corpus of 2 trillion tokens sourced from publicly available materials. 

For each speech sequence, a transcript of the audio clip is tokenised and processed by the \llama2 model. The \llama2 model extracts a sequence of embeddings, with each embedding corresponding to the respective input token. For each word in the utterance, we assign an output embedding and execute frame-wise alignment to ensure that each embedding is synchronized with its corresponding motion frame timing at 30fps. The process generates text-embedding vectors $T$ of shape $(N, 4096)$. 

\begin{figure}[tb]
\centering
  \includegraphics[width=\linewidth]{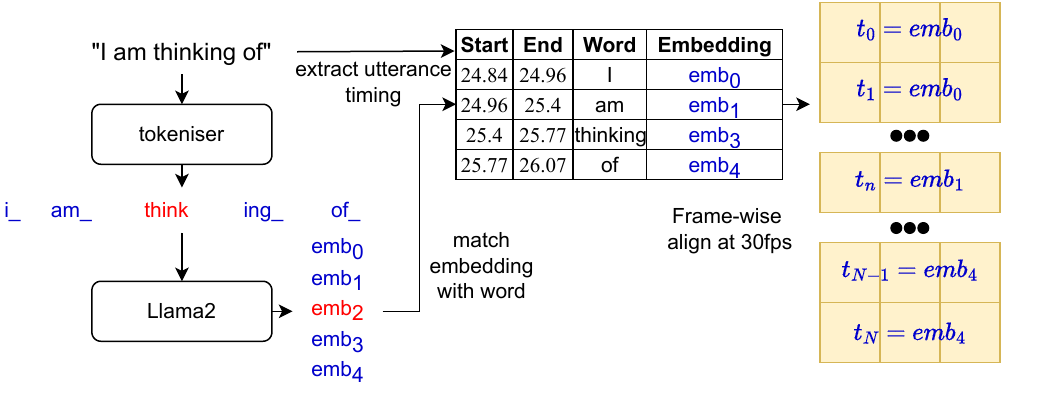}
  \caption{Extracting text features using \llama2. The text is BPE-tokenised, and a \llama2 embedding is computed for each token. These embeddings are aligned with audio at 30fps by repeating frames as necessary.} 
  \label{fig:text}
\end{figure}

Alignment is achieved by repeating text embeddings as needed to synchronise with the audio timing. In instances where a word spans multiple frames, the vector is duplicated for the corresponding number of frames, and a zero-value vector is employed when no word is spoken at a specific frame.
Figure \ref{fig:text} provides an overview of the alignment process.

The input utterance is tokenised using a \ac{bpe} method, meaning a single word may be broken into multiple constituent parts. For example, the word ``thinking'' will be divided into two tokens, ``think'' and ``ing''. 
In such cases, only the embedding for the last token is retained, and the embeddings for the preceding parts are discarded. 
For example, the embedding associated with ``ing'' is used rather than ``think''. This is common practice when using \ac{llm}s as the final embedding is expected to encapsulate information about preceding tokens.

\subsubsection{Speaker style}
For each utterance, a speaker label is additionally provided as input. 
This is a unique ID per speaker which is passed through a learned embedding layer.
The trainable weights of this layer ensure that speakers with similar gesture styles are positioned closely in the latent embedding space, while speakers with distinct gesturing styles are situated further apart. We use an 8-dimensional embedding to generate speaker vectors $S$ with a shape of $(N, 8)$.

\subsection{Body Pose Representation}
The body pose at time $n$ is defined as:
\begin{equation}
\mathbf{y_n} = [x_n, y_n, z_n, r_{j, 1, n}, ... , r_{j, 6,N}]
\end{equation}
where $x, y, z$ denote the global skeleton position and $r_{j,1:6, n}$ form rotations for each joint $j$ in the 6D rotation representation presented by \cite{Zhou_2019_CVPR}. These values are standardised by subtracting the mean and dividing by the standard deviation computed from the training data.

\subsection{Model Architecture}
In this study, our primary objective is to evaluate the impact of \ac{llm} features on the animation of co-speech gestures.
To accurately measure this effect, we employ an established model and training method. Specifically, we adopt a model based on the Cross-Attentive Transformer-XL, which demonstrated effectiveness in the \ac{genea} challenge 2023 \cite{windlegenea23}.
This approach is built on the Transformer-XL model architecture \cite{dai2019transformer} which uses segment-level recurrence with state reuse and a learned positional encoding scheme to ensure temporally cohesive boundaries between segments.
Windle et al. extend this architecture using cross-attention to incorporate the second speaker's speech into the prediction when used in a dyadic setting.
Notably, this architecture delivers high-quality results without the need for more involved training techniques such as diffusion.

Either a single modality or a combination of features are used to form the input feature vectors $X \in \{X_a, X_b, X_+, X_\times\}$. Please refer to Section~\ref{sec:inputfeatures} for more details on the construction of this matrix. 
We train our model on dyadic conversation between a main-agent and interlocutor.
Specifically, we predict the main-agent's gesturing conditioned on both main-agent and interlocutor speech.
Consequently, we compute a set of input features for each speaker, $X^{ma}$ and $X^{in}$, and a set of target poses for the main-agent, $Y$.
These extracted features are segmented into non-overlapping segments of length $W$ frames.

\begin{figure}[tb]
\centering
  \includegraphics[width=\linewidth]{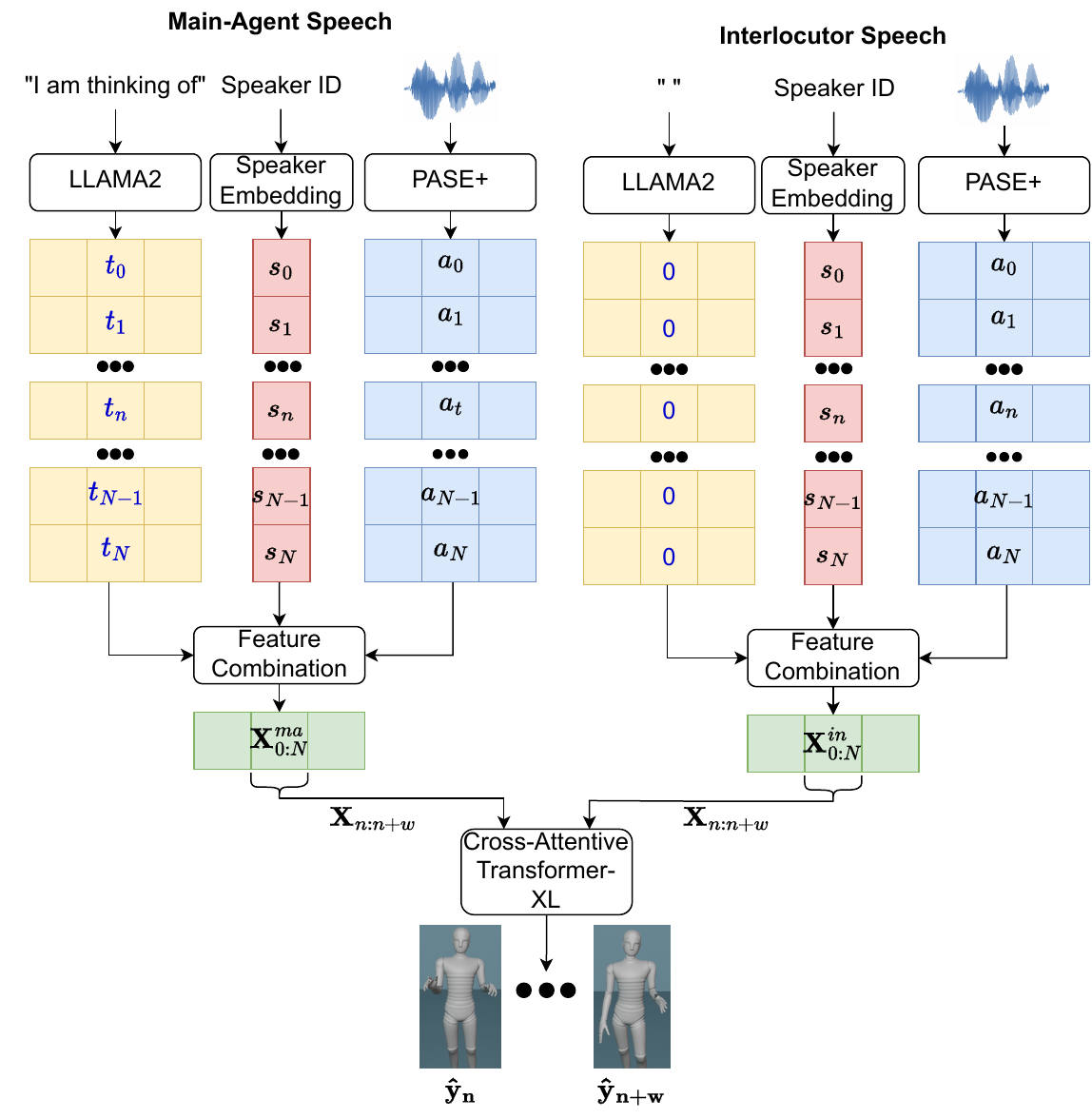}
  \caption{Overview of \llanimation \space method. Our model takes \llama2 features as input, along with a speaker embedding and optional \ac{pase} features that encode the speech of a main-agent and an interlocutor. The features are combined and processed through a cross-attentive Transformer-XL model that produces gesture animation for the main-agent.}
  \label{fig:overview}
\end{figure}

Given an input feature vector $X$ of length $W$, the Transformer-XL predicts $\hat{Y}$ of length $W$ using a sliding window technique with no overlap.
Consequently, for a speech sequence of length $N$, our model is invoked $\lceil\frac{N}{W}\rceil$ times.
Figure \ref{fig:overview} shows an overview of this approach.

\subsection{Training Procedure}
We follow the same training methodology as in \cite{windlegenea23}
and include the same geometric and temporal constraints in the loss function. 
The loss function $L_c$ comprises multiple terms including a $L_1$ loss on the rotations ($L_r$), positions ($L_p$), velocity ($L_v$), acceleration ($L_a$) and kinetic energy ($L_{v^2}$) of each joint. 

All training parameters were kept the same as in Windle et al. However, an additional two self-attention layers were included in the Cross Attentive Transformer-XL. These additional layers were chosen based on validation loss values and the quality of the predicted validation sequences, as observed by our team.
We train our models for 1300 epochs using the AdamW optimiser \cite{loshchilov2017decoupled}. 

\subsection{Smoothing}
The raw model output can contain low levels of high-frequency noise. Following other work on motion synthesis \cite{zhang2023diffmotion, zhou2022gesturemaster}, we apply a Savitzky-Golay Smoothing filter to mitigate this. We use a window length of 9 and polynomial order of 2. The small window size and low polynomial means this filter provides a very small amount of localised smoothing while retaining accurate beat gestures.
\section{Experimental Setup}
Four distinct models are trained, each with a different set of features: 1) PASE+: An audio-only model, 2) \llanimation: A \llama2 text-only model, 3) \llanimation-$+$: A \llama2 and PASE+ concatenated model and 4) \llanimation-$\times$: A \llama2 and PASE+ cross-attention model. In this section we describe our data and details on the model configurations.

\subsection{Data}\label{sec:data}
The data used in this study is from the \ac{genea} challenge 2023 \cite{kucherenko2023genea}. This dataset is derived from the Talking With Hands dataset \cite{lee2019talking}, containing dyadic conversations between a main-agent and interlocutor. It comprises high-quality 30fps motion capture data in Biovision Hierarchical (BVH) format. The dataset includes both speech audio and text transcripts derived from both speakers in the conversations. 

The dataset is divided into three splits: 1) train, 2) validation, and 3) test. The validation set is employed for model tuning and refinement, while the test set is exclusively reserved for evaluation.

\subsection{Feature Combinations}\label{sec:inputfeatures}
Our experiments use audio and text modalities in isolation and additionally investigate two approaches for combining the two modalities: 1) post-extraction concatenation and 2) cross-attention, respectively referred to as \llanimation-$+$ and \llanimation-$\times$.

\subsubsection{Single Modalities}
To use each modality individually, we concatenate the speaker $S$ matrices with the audio $A$ or text $T$ along the feature dimension to form $X_a$ and $X_t$, respectively. The concatenated matrix is then passed through a linear layer, giving:
\begin{equation}
\begin{aligned}
X_a = W_a (A,S)^\top + \mathbf{b}_a \\
X_t = W_t (T,S)^\top + \mathbf{b}_t
\end{aligned}
\label{eq:singlemodalities}
\end{equation}
where $W_a$, $W_t$, $\mathbf{b}_a$ and $\mathbf{b}_t$ are learned parameters. $X_a$ and $X_t$ are used as inputs for training the single modality audio and text-based models respectively.

\subsubsection{Concatenation}
To combine modalities we concatenate $A$, $T$ and $S$ matrices along the feature dimension. The concatenated matrix is then passed through a linear layer, giving:
\begin{equation}
X_+ = W (A,T,S)^\top + \mathbf{b}
\end{equation}
where $W$ and $\mathbf{b}$ are learned parameters. This results in $X_+$ which are the concatenated audio and text features and serve as the input to  \llanimation-$+$. 

\subsubsection{Cross-attention}
Additionally, we experiment with using cross-attention for combining audio and text features. Cross-attention has been shown to be an effective method of combining modalities, as evidenced in \cite{ng2024audio}. 
In this approach, we first concatenate the style embedding to both audio and text features. We then linearly project the two concatenated matrices into the same feature dimension size, $d$, following Equation~\ref{eq:singlemodalities}.
We perform cross-attention on the feature dimension, such that the projected audio features, $X_a$, serve as the query, while the projected text features, $X_t$ are set as the key and value \cite{vaswani2017attention}:
\begin{equation}
X_\times = \mathrm{softmax}(\frac{X_aX_t^\top}{\sqrt{d}})X_t
\end{equation}
giving the cross attention combined audio and text features $X_\times$ which are used as input for training \llanimation-$\times$.
\section{Evaluation}
We present an evaluation into the efficacy of \llama2 features for gesture generation, in isolation and in combination with audio PASE+ features. We present our observations and report the associated performance metrics. Finally, we describe a user study that measures the differences in perceived quality. 

\subsection{Observations}
We observe noticeable differences between the animation produced by the PASE+-based model and the {\llanimation} method. The PASE+ version primarily generates beat gestures, whereas the {\llanimation} model exhibits more varied motions, encompassing both beat and semantic gestures. The animation from {\llanimation} appears to be more expansive and confident. 
Video examples and comparisons showing this effect can be seen in the supplementary material\footnote{\href{https://youtu.be/jBXpWocXvZ8}{https://youtu.be/jBXpWocXvZ8}}.

\subsubsection{Beat Gestures}
Beat gestures are characterised by simple and fast movements of the hands, serving to emphasise prominent aspects of the speech~\cite{bosker2021beat}. These gestures have a close relationship with the timing of prosodic activity, such as acoustic energy and pitch \cite{windle2022arm, pouw2020energy}.  
Given that these prosodic features can be directly derived from the audio signal, an audio-based model can be very effective at generating beat gestures. 
A beat gesture is not necessarily expected for every audio beat, but when performed, it is likely to be well-timed with the audio beats.

\begin{figure}[tb]
  \centering
\includegraphics[width=.75\linewidth, trim=0 80 0 0,clip]{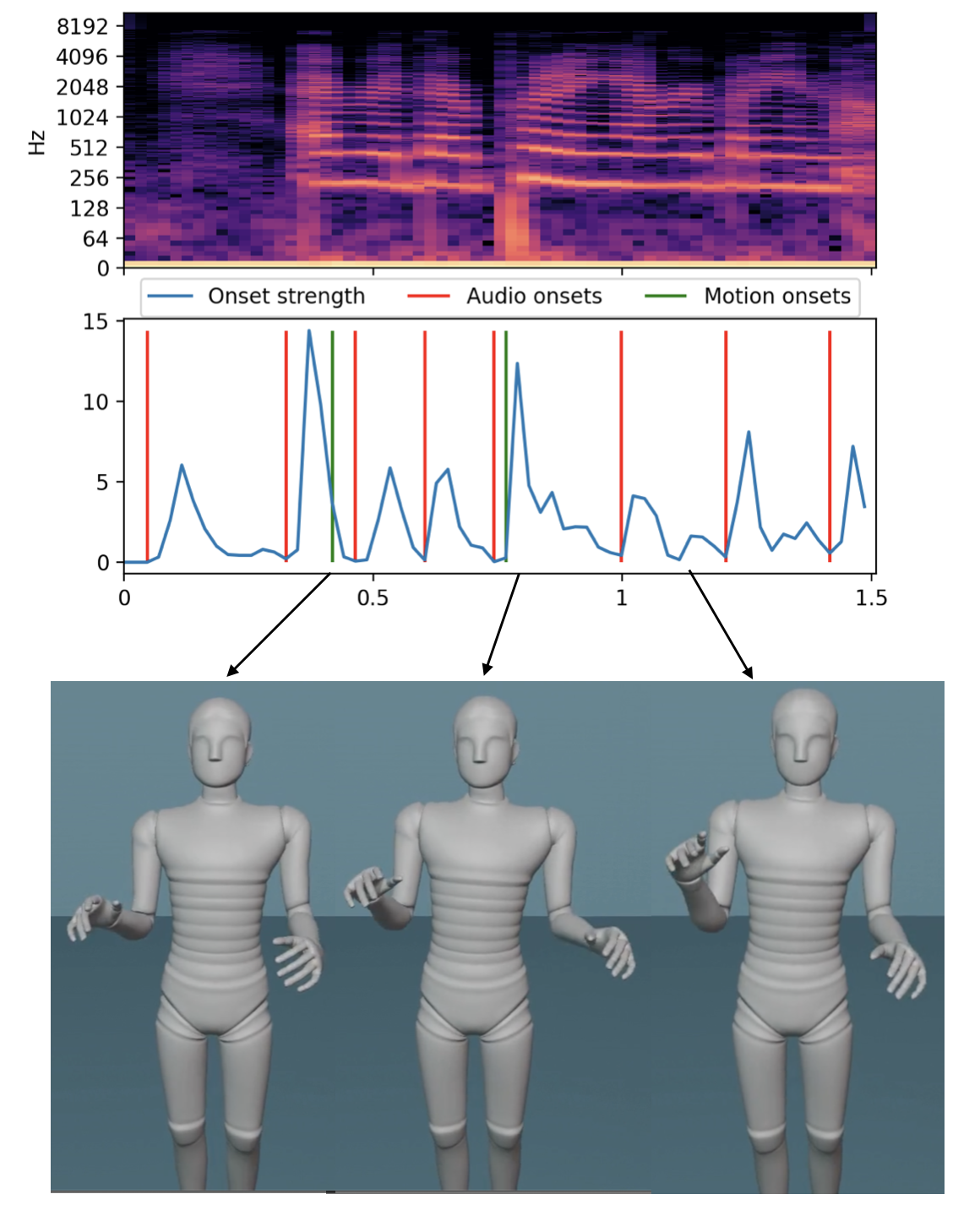}
  \caption{Generated gestures for given audio beats using \llanimation \space method. Using a 1.5s audio clip from the test dataset, we show the audio spectrogram, as well as aligned audio beat onsets and their corresponding onset strengths, as well as motion gesture onset detection of the left wrist using the method of beat detection defined in \cite{liu2022beat}. The speaker moves their left hand from right to left and back again as the syllables are stressed.} 
  \label{fig:timing}
\end{figure}

Using the motion and audio beat extraction method defined in the beat align score calculation proposed by \cite{liu2022beat}, we can visualise the onset of audio beats and motion gestures over time.
Remarkably, we observed that \llanimation \space with \llama2 and no audio features consistently executes beat gestures in synchronisation with audio beats, despite lacking explicit energy or pitch information.
Figure \ref{fig:timing} shows a 1.5-second clip with the left wrist motion onsets in green and audio beat onsets in red. A speaker can be seen swiftly moving their left hand from left to right in time with audio beats and returning close to their original pose. 

Although we temporally align the \llama2 embeddings providing the model with awareness of word timings, there is no explicit knowledge of syllable-level timing. 
Further investigation is needed; however, it is plausible that training with \llama2 embeddings may effectively encode information regarding the presence of lexically stressed syllables in context within words.

\subsubsection{Semantic gestures}
Semantic gestures are often directly linked to speech content, such as mimicking an action or nodding the head in agreement. In our observations, the \llanimation \space method demonstrated superior performance compared to the audio PASE+-based model in generating these types of gestures.

In a test sequence where a speaker is describing the act of eating a crab, the \llanimation \space gestures exhibit more activity compared to the PASE+ version, particularly when the speaker uses their hands to illustrate actions. This is exemplified when the hands mimicked sticking a fork in a crab for consumption in time with the verbal description. This sequence can be seen in Figure \ref{fig:teaser} and the supplementary video.

\begin{figure}[tb]
\centering
  \includegraphics[width=.8\linewidth, trim=500 650 680 280,clip]{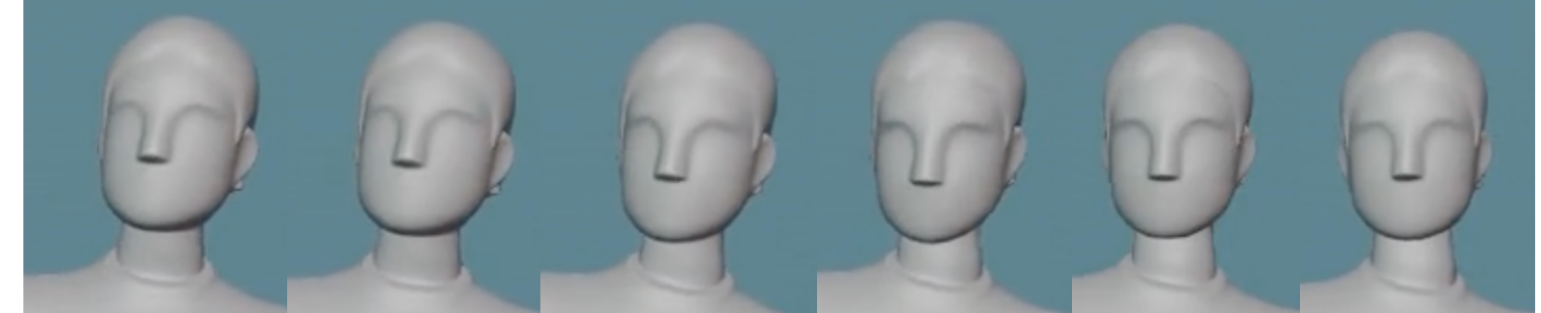}
  \caption{Example nod motion temporally aligned with the word ``yes'' being spoken. from a test sequence generated using the \llanimation \space } 
  \label{fig:head}
\end{figure}

\llanimation \space demonstrates the capacity to adequately encode agreeableness. For example, Figure \ref{fig:head} shows a predicted test sequence where the speaker can be seen nodding along with the word yes. 

\subsubsection{Laughter}
\begin{figure}[tb]
\centering
  \includegraphics[width=.7\linewidth]{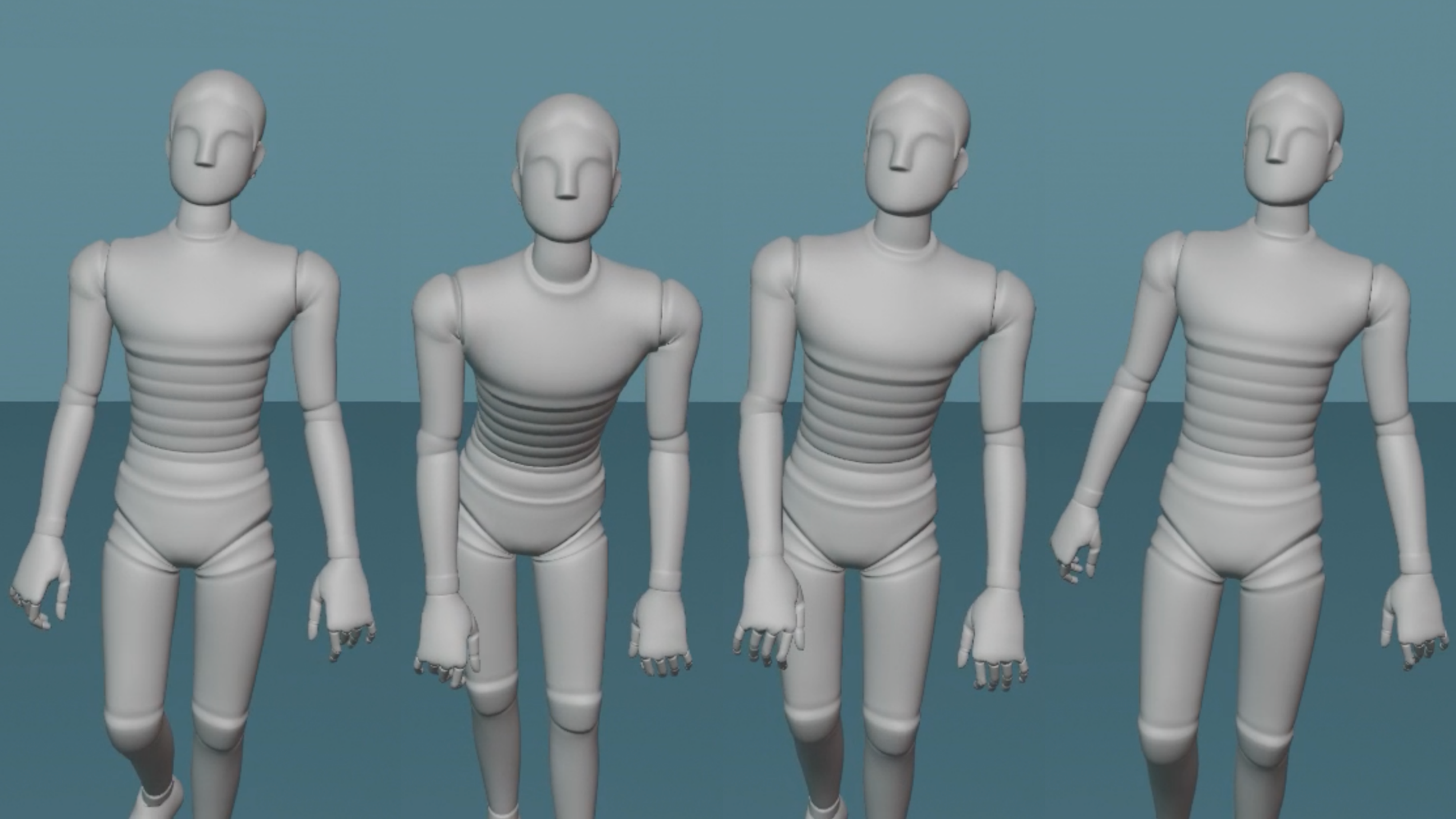}
  \caption{Example laughter sequence generated using the \llanimation \space method} 
  \label{fig:laugh}
\end{figure}

During the transcription process of the \ac{genea} dataset, laughter without speech was denoted using ``\#\#\#''. This representation was directly input to the \llama2 model for feature extraction. Although the generated embedding would not encode any semantic meaning, our model learns to associate these tokens to laughter. The \llanimation \space method captures moments of laughter as illustrated in Figure \ref{fig:laugh}, where the character partially creases over. This specific behaviour is not observed in the gesture animation produced by the PASE+-based model .


\subsection{Performance Metrics}
Evaluating the objective performance of gesture generation poses a challenging research question, primarily due to the many-to-many ambiguous relationship between speech and gesture. No single metric has been developed that correlates with human perception. However a \emph{combination} of metrics can be used as a means to somewhat evaluate the quality of generated gesture. Frèchet Gesture Distance (FGD) \cite{yoon2020speech,bhattacharya2021speech2affectivegestures, ng2024audio}, Frèchet Kinetic Distance (FD$_k$) \cite{ng2024audio} and Beat Alignment (BA) \cite{li2021ai, liu2022beat} are useful metrics for this task. These metrics are indicative of static and dynamic appropriateness, and the alignment of motion to speech \cite{alexanderson2023listen, liu2022beat, yoon2020speech}.
Frèchet Gesture Distance is a measure based on the Frèchet Inception Distance (FID) \cite{heusel2017gans}, which is commonly used for evaluating generative models. 
A pre-trained autoencoder extracts domain-specific latent features from both ground truth and predicted motion. 
The FGD score is a Frèchet distance between the two multivariate Gaussian distributions of these features in latent space.
This measures similarity between the generated and ground truth poses but does not necessarily indicate how well the generated examples temporally align with the audio.

Frèchet Kinetic Distance is similar, however, there is no auto-encoding process. Instead, the first derivative of each joint is used to determine the distribution of velocities for both the ground truth and predicted motion. FD$_k$ is the Frèchet Distance between these two distributions.

Beat Alignment has been adapted from music synthesis \cite{li2021ai} to work with gesture generation \cite{liu2022beat}.
Using a chamfer distance between audio and gesture beats, this gives a synchrony measure between the two. Beats are detected using the root mean square onset of the audio and a motion beat is identified by the local minimums of the velocity.

\subsubsection{Results}

The measures presented in Table \ref{tab:objective-audio} indicate that the FGD and FD$_k$ scores are consistently lower for all \llama2-based  models than for the model trained on PASE+ features. This suggests that the motion generated by \llanimation \space may be closer to ground truth, with \llanimation-$+$ showing the most realistic motion.
The BA score suggests that the audio features are the most timely, however, the differences between this and the \llanimation \space methods are minimal. Notably, the method with no audio features is competitive in FGD and BA scores.

\begin{table}[tb]
\begin{tabular}{l|ccc}
\textbf{Model} & $FGD\downarrow$ & $FD_{k}\downarrow$ & BA$\uparrow$ \\ \hline
PASE+            & 79.90        &     34.37                & \textbf{0.871}       \\
\llanimation             & 61.86       &            24.23         & 0.855       \\
\llanimation-$+$     &   \textbf{47.56}        &     \textbf{23.79}                &     0.869        \\
\llanimation-$\times$     & 66.87        &           25.70          & 0.865      
\end{tabular}
\caption{Frèchet Gesture Distance (FGD)
,  Frèchet Kinetic Distance ($FD_{k}$) and Beat Alignment (BA)
scores for each system calculated with respect to the ground truth test dataset.}
\label{tab:objective-audio}
\vspace*{-0.5cm}

\end{table}

\subsection{User Study}\label{sec:userstudy}
We present a user study to further evaluate perceived human likeness and appropriateness of the animations from the PASE+-based model compared with the \llama2-based {\llanimation} method. 
Participants were hired through the Prolific\footnote{https://www.prolific.co/} platform with 50 participants in each experiment after removing any participants that failed attention checks.
Participants were filtered to be fluent in English.
For this study, we used a similar methodology to \cite{alexanderson2023listen} and the \ac{genea} Challenge 2023 \cite{kucherenko2023genea}.

All test sequences for each method were rendered on the same virtual avatar released by \cite{kucherenko2023genea}, as shown in Figure \ref{fig:teaser}. 
We use the exact clip timings from the \ac{genea} Challenge, comprising 41 clips with an average duration of 10 seconds each. During evaluation, users exclusively heard the audio of the main-agent being animated. 

In our pairwise system comparison, participants were presented with two side-by-side videos generated for the same audio but with different systems.
To mitigate bias, we randomise the question order and randomly swap the side of the screen that each condition is shown.

The question for all studies was posed as ``Which character’s motion do you prefer, taking into account both how natural-looking the motion is and how well it matches the speech rhythm and intonation?''.
The participants were asked to choose from the options \{\textbf{Clear preference for \textit{left}}, \textbf{Slight preference for \textit{left}}, \textbf{No Preference}, \textbf{Slight preference for \textit{right}} and \textbf{Clear preference for \textit{right}}\}.
The scoring methodology uses a merit system \cite{parizet2005comparison} where an answer is given a value of 2, 1 or 0 for clear preference, slight preference and no preference, respectively. 
Preference testing allows a win rate calculation where a win is assigned when there is an identified preference for a system, not including ties.
A one-way ANOVA test with a post-hoc Tukey test was subsequently used for significance testing.

\begin{table*}[tb]
\begin{tabular}{l|c|cc|cc|cc|cc} 
\hline                                                            & & \multicolumn{2}{c|}{vs PASE+} &  \multicolumn{2}{c|}{vs \llanimation} &  \multicolumn{2}{c|}{vs \llanimation-$+$}  &  \multicolumn{2}{c}{vs \llanimation-$\times$}             \\
    & Merit Score            & Win Rate & Tie Rate & Win Rate & Tie Rate & Win Rate & Tie Rate & Win Rate & Tie Rate \\
\hline 
PASE+         & 0.37$\pm$0.05          & -   & - & 25.4\%   & 11.4\% & 24.6\% & 14.8\% & 28.4\% & 14.8\% \\
\llanimation        & 0.68$\pm$0.06 & 63.3\%        & 11.4\%  & -   & - & 38.6\% & 22.3\% & 44.3\% & 14.8\%     \\
\llanimation-$+$         & \textbf{0.69}$\pm$0.06             & 61.0\% & 14.8\% & 39.0\%   & 22.3\% & - & - & 43.2\% & 20.8\%  \\
\llanimation-$\times$ & 0.64$\pm$0.06          & 56.8\%  & 14.8\% & 40.9\%   & 14.8\% & 36.0\% & 20.8\% & - & -  \\
\hline
\end{tabular}
 \caption{User study results. Merit scores \cite{parizet2005comparison} with 95\% confidence intervals, win and tie rates for each comparison.
}
\label{tab:subjective-audio}
\end{table*}

\begin{table*}[tb]
\begin{tabular}{l|c|cc|cc|cc|cc} 
\hline                             
         & & \multicolumn{2}{c|}{vs GT}   &   \multicolumn{2}{c|}{vs \llanimation} & \multicolumn{2}{c|}{vs \llanimation-$+$} & \multicolumn{2}{c}{vs CSMP-Diff}  
\\
    & Merit Score            & Win Rate & Tie Rate  & Win Rate & Tie Rate & Win Rate & Tie Rate & Win Rate & Tie Rate \\
\hline 
GT         & 1.16$\pm$0.05  & - & -  & 78.6\% & 8.9\% & 74.8\% & 8.9\% & 68.9\%   & 11.1\% \\
\hdashline
\llanimation        & 0.34$\pm$0.04    & 12.5\% & 8.9\%  & - & - & 34.2\% & 33.6\% & 31.4\%        & 16.1\% \\
\llanimation-$+$         & 0.36$\pm$0.04            & 16.4\% & 8.9\%  & 32.2\% & 33.6\% & - & - & 35.6\%   & 14.4\% \\
CSMP-Diff         & \textbf{0.58}$\pm$0.05           & 20\% & 11.1\%  & 52.5\% & 16.1\% & 50.0\% & 14.4\% & -  & -\\
\hline
\end{tabular}
 \caption{User study results. Merit scores \cite{parizet2005comparison} with 95\% confidence intervals, win and tie rates for each comparison.
}
 \label{tab:subjective-gt}
\end{table*}

\subsubsection{Results}
Table \ref{tab:subjective-audio} summarises the results of the user study. 
These findings validate the objective measure scores in that all \llanimation-based models outperform the PASE+ audio-only method. According to the merit score, all \llanimation \space methods were significantly preferred over the PASE+ approach ($p<0.001$).
Win and tie rates show that \llanimation \space methods win or are tied with PASE+ most of the time. Surprisingly, the highest win rate is recorded by the \llanimation \space method with no PASE+ features included, suggesting that using text as a sole input is sufficient to generate plausible speech gesturing, and that audio features are somewhat redundant in our model

Between each \llanimation \space method, there is no statistically significant difference in merit scores. We examine the win and tie rates against \llanimation \space to determine if adding PASE+ features will provide additional preference. We can see from these rates that the choice between \llanimation \space settings is almost tied to wins and losses. \llanimation-$+$ wins 1.9\% less than \llanimation-$\times$; however, the tie rate is higher and therefore loses less than \llanimation-$\times$. 

This initial study concludes that \llama2 features are powerful at encoding information useful to gesture generation and can produce more realistic-looking gestures than a model trained on audio input. Combining modalities also does not make a significant difference, although the concatenation of features performs slightly better than the cross-attention regarding merit scores and win/tie rates.

\begin{table}[tb]
\begin{tabular}{l|lll}
\textbf{Model} & $FGD\downarrow$ & $FD_{k}\downarrow$ & BA$\uparrow$ \\ \hline
\llanimation             & 61.86       &            24.23         & 0.855       \\
\llanimation-$+$     &   47.56        &     23.79                &     \textbf{0.869}        \\
CSMP-Diff     & \textbf{30.620}        &           \textbf{12.61}          & 0.866      
\end{tabular}
\caption{Frèchet Gesture Distance (FGD) \cite{yoon2020speech}, Frèchet Kinetic Distance ($FD_{k}$) and Beat Alignment (BA) \cite{liu2022beat} scores for each system calculated with respect to the ground truth test dataset.}
\label{tab:objective-gt}
\end{table}
\section{Comparison Against Other Systems}

Our previous study has shown that we achieve a significant performance improvement achieved by integrating \ac{llm} features into gesture-generation models. We now perform additional experiments to compare our best performing \llanimation \space and \llanimation-$+$ approaches against both ground truth and the current state-of-the-art method. This broader evaluation aims to assess performance across the field.

We compare against the state-of-the-art \ac{csmp} method \cite{deichler2023diffusion}, which achieved the highest human-likeness and speech appropriateness rating among the entries to the 2023 \ac{genea} challenge.

Objective performance metrics are shown in Table \ref{tab:objective-gt}. \ac{csmp} performs better in FGD and FD$_k$ scores. We find minimal differences to the BA score, with \llanimation \space marginally outperforming \ac{csmp}.

We repeated the user study following the protocol as described Section~\ref{sec:userstudy}, and the results are summarised in Table \ref{tab:subjective-gt}. 
 In terms of merit score, the ground truth was perceived as significantly better than any other method ($p<0.001$), underscoring the current challenge in consistently generating human-realistic gesturing. CSMP-Diff was considered superior to both \llanimation \space methods ($p<0.001$). 
  Despite this difference, when examining the win rates against \ac{csmp}, we find that the \llanimation \space method wins 31.4\% of the time and ties 16.1\%. Meanwhile, our \llanimation-$+$ method won 35.6\% of the time and ties 14.4\%. In each case, \llanimation \space and \llanimation-$+$ are rated as good or better than \ac{csmp} 47.5\% and 50\% of the time, respectively.
 
\ac{csmp} incorporates both diffusion and contrastive speech and motion pre-training, representing two advanced and complex techniques. Despite these sophisticated methods, our evidence indicates that \llanimation, even in the absence of any audio input, can perform as well as or better than \ac{csmp} nearly half the time. This suggests that \llama2 features serve as incredibly valuable feature encodings for gesture animation.

\section{Conclusion}
We have explored the use of \llama2 features for speech-to-gesture generation in our proposed {\llanimation} method. 
With the use of \llama2 features we were able to generate well timed and contextually rich gestures even without the inclusion of any audio feature embedding.
We explored the use of combining both audio and text modalities through concatenation and cross-attention and found that there was no significant difference in the inclusion of \ac{pase} features when compared to using \llama2 features in isolation. 
We have demonstrated the performance improvements when incorporating the \llama2 features into a gesture-generation model through both objective and subjective measures.
Given this finding, we believe that human speech related gesture animation is heavily related to the semantic encoding that is present in the \llama2 embeddings, and that these embeddings additionally capture a notion of prosody from the language context. This is a somewhat surprising finding, and a result we think can have great practical impact on future content-aware animation systems.

We additionally compared our {\llanimation} approach to ground truth as well as the state-of-the-art \ac{csmp} approach. The evaluation revealed that both \llanimation \space and \ac{csmp} have areas where improvement is possible as they are unmatched against ground truth. While \ac{csmp} remains state-of-the-art, it is a complex model and our simpler alternative was rated as good or better than it 50\% of the time. We predict that integrating \ac{llm} features into state-of-the-art systems will be a step towards bridging the gap between machine-generated and natural gesturing.

\subsection{Future Work}
While we show that the use of \ac{llm} features can be powerful for generating contextually and semantically correct gestures, more work is required to get performance closer to the ground truth. 
We use the 7-billion parameter release of \llama2 in this work due to hardware constraints. With more resources, the larger 70-billion parameter could be utilised, which may produce more nuanced and varied gesturing. 
We do not fine-tune the \llama2 model for our domain. \ac{llm}s are known to perform well with prompting and in-context learning \cite{yang2023harnessing} to fine-tune the model. We therefore foresee many opportunities for further performance gain.


\bibliographystyle{ACM-Reference-Format}

\bibliography{sample-base}

\end{document}